\documentclass{article}
\usepackage{spconf,amsmath,graphicx}
\usepackage{xcolor}

\usepackage{enumitem}
\usepackage{soul}
\setlist{nosep, leftmargin=14pt}

\usepackage{mwe} 

\usepackage{booktabs}
\usepackage{bm}
\usepackage[super]{nth}
\usepackage{amssymb}
\usepackage{hyperref}
\usepackage[utf8]{inputenc}
\usepackage{placeins}
\usepackage{float}
\usepackage{afterpage}
\usepackage{graphicx}
\usepackage{caption}

\DeclareCaptionLabelFormat{onlynumber}{#2}

\title{Predicting Future States with Spatial Point Processes in Single Molecule Resolution Spatial Transcriptomics}

\name{
     Biraaj Rout$^{\star, 1, 4}$ \qquad
      Priyanshi Borad $^{\star, 2}$ \qquad
 Parisa Boodaghi Malidarreh$^{\star, 1, 4}$ 
}

\secondlinename{
     Mohammad Sadegh Nasr $^{1, 4}$ \qquad
    Jillur Rahman Saurav$^{ 1, 4}$ \qquad
    Kelli Fenelon$^{2}$ \qquad
}

\thirdlinename{
    Jai Prakash Veerla$^{1, 4}$ \qquad
    Jacob M. Luber $^{\dagger,1, 3, 4}$ \qquad
    Theodora Koromila$^{\dagger, 2,5}$
}

\address{
    $^{1}$ Department of Computer Science and Engineering, University of Texas at Arlington\\
    $^{2}$ Department of Biology, University of Texas at Arlington\\
    $^{3}$ Department of Bioengineering, University of Texas at Arlington\\
    $^{4}$ Multi-Interprofessional Center for Health Informatics, University of Texas at Arlington\\
    $^{5}$ School of Biology, Aristotle University of Thessaloniki
}

\begin{document}
\maketitle
    \def\thefootnote{$\star$}\footnotetext{Equal contribution.}
    \def\thefootnote{$\dagger$}\footnotetext{Responsible authors. Email: \texttt{jacob.luber@uta.edu}, \texttt{theodora.koromila@uta.edu}}
    
    \begin{abstract}
In this paper, we introduce a pipeline based on XGBoost (eXtreme Gradient Boosting) to predict the future distribution of cells that are expressed by the $sog$ gene (active cells){ in both the Anterior to posterior (AP) and the Dorsal to Ventral (DV) axis} of the \textit{Drosophila} in embryogenesis process. This method provides insights about how cells and living organisms control gene expression in super resolution whole embryo spatial transcriptomics imaging at sub cellular, single molecule level. An XGBoost model was used to predict the next stage active distribution based on the previous one. To achieve this goal, we leveraged temporally resolved, spatial point processes by including Ripley's K-function in conjunction with the cell's state in each stage of embryogenesis, and found average predictive accuracy of active cell distribution. This tool is analogous to RNA Velocity for spatially resolved developmental biology, from one data point we can predict future spatially resolved gene expression using features from the spatial point processes.       \end{abstract}


    \section{Introduction}
\label{sec:intro}
Recent technological advances have made it possible  to capture high resolution images from embryogenesis processes that help researchers to study gene expression patterns.\cite{koromila2019distinct, dunipace2013autoregulatory}.
One of the major challenges of the modern genomics era is better understand how gene expression is regulated to support spatiotemporal outputs that change over the course of development. The early \textit{Drosophila} embryo has served as a paradigm for how enhancers control patterning and has demonstrated that the patterning process is complex and dynamic. It is known that multiple, transiently acting enhancers function sequentially to regulate dynamic changes in gene expression outputs \cite{dunipace2013autoregulatory,long2016ever, perry2012precision}, whereas other genes are controlled by enhancers that act over a longer period and support changing spatial outputs over time. For example, expression of the gene short gastrulation  (\textit{sog}) is driven by at least two co-acting enhancers that support temporally dynamic expression. Live imaging experiments enable the potential to analyze gene expression dynamics with increased temporal resolution and linear quantification. However, genetic and live imaging techniques have outpaced analysis approaches to harvest the bountiful information contained within real-time movies of transcriptional dynamics with modern methods confined to static parameter cell and transcript tracking methods \cite{koromila2019distinct, lim2018visualization,birnie2023precisely}. To assess these mutant enhancer phenotypes systematically, we developed a quantitative approach to measure the spatiotemporal outputs of enhancer-driven MS2 reporter constructs as captured by \textit{in vivo} imaging to provide information about the timing, levels, and spatial domains of expression. Using transgenic fly lines, we conducted live imaging to visualize GFP signal associated with MS2 stem-loop reporter sequence binding MCP-GFP, enabling dynamic tracking of RNA localization and expression in real time. This MS2 cassette contains 24 repeats of a DNA sequence that produces an RNA stem loop when transcribed. The stem-loop structure is specifically bound by the phage MS2 coat protein (MCP). MCP fused to GFP binds to MS2-containing transcripts (i.e., \textit{sog\_Distal}.MS2) producing a strong green signal within the nuclei of \textit{Drosophila} embryos at sites of nascent transcript production. In this system, nuclear GFP fluorescence is observed as a single dot per nucleus in heterozygous individuals, corresponding to nascent transcription from a single copy of the MS2-containing reporter transgene integrated into the genome. Furthermore, the nuclear periphery is marked by a fusion of RFP to nuclear pore protein (Nup-RFP) \cite{lucas2013live}. We optimized the imaging protocol to provide spatial information across the entire dorsal-ventral (DV) axis of embryos with the fastest temporal resolution that also retains embryo viability. In brief, embryos were imaged on a Zeiss LSM 900 continuously over the course of 2hr at an interval of ~30s per scan (twice as fast compared to previous studies). Importantly, this imaging protocol is not photo-toxic to embryos. Because spatial outputs likely change in time across the embryo for many gene expression patterns, we developed an image processing approach to collect detailed information in both time and space by capturing one lateral half of the embryos. With this qualified imaging dataset, our goal was to predict the distribution of active cells in each stage of embryonic development as the blastula transitions into gastrulation. Building on existing methodologies for predicting temporal variables; authors in \cite{la2018rna} introduced the concept of RNA velocity, which is defined as the time derivative of gene expression, offering a novel approach for inferring dynamic changes in gene activity over time. This concept allows for the estimation of future states of individual cells in standard scRNA-seq protocols. In \cite{dayao2023deriving}, authors proposed a method to capture spatial proteomics data to map cell states in order to predict cancer patient survival. They utilized the Ripley's K-function for capturing spatial features which inspired us in our proposed pipeline. 
We developed a feature extraction method and analysis pipeline that can be used to predict the future distribution of cells in which the \textit{sogD} gene is expressed.

    \section{methods}
\label{sec:intro}

\subsection{Experimental set-up for embryo collection }
\label{subsec: intro}
Virgin females expressing MCP-GFP (green) and Nup-RFP (red) maternally were crossed with males carrying either the $sog\_Distal$ eve2 promoter-MS2.yellow-attB [Broadly expressed repressors integrate patterning across orthogonal axes in embryos] or $sogD\_ {\Delta Su(H)}$ eve2 promoter-MS2.yellow-attB \cite{koromila2019distinct}. Embryos were precisely timed and collected during nuclear cycles 10–11. After collection, embryos were carefully dechorionated and mounted in Halocarbon 27 between a slide and coverslip, with spacing achieved using double-sided tape, as previously described \cite{fenelon2024h}.

\subsection{Live Imaging}
\label{subsec: intro}
Embryos were collected on apple agar plates for 1 hour, rested for 30 minutes at room temperature, and manually dechorionated. They were mounted between a slide and coverslip using heptane-dissolved adhesive and immersed in Halocarbon 27 oil. Imaging was performed on a Zeiss LSM 900 Airyscan 2 (Zeiss, Oberkochen, Germany) during stages leading into gastrulation, with broad-view and super-resolution movies captured using a 40× water oil immersion objective. Images were acquired at varying resolutions and intervals, as described \cite{fenelon2024h}.

\subsection{Data Pre-processing}
\label{subsec: intro}
We conduct pre-processing, feature extraction, training, and testing Fig.\ref{fig:2}. Both the training and testing phases incorporate identical pre-processing and feature extraction steps. The videos shows real time images  from embryonic development, which were manually given stage development labels: NC 13, NC 14 A, NC 14 B, NC 14 C, NC 14 D. In the pre-processing step, we used a generalist, deep learning-based segmentation method called Cellpose, which can precisely segment cells in each frame of the embryo development. Active cells were identified based on prevalance of green pixels indicative of gene expression within the cell, and the active mask underwent feature extraction. During this stage, the masked images underwent a gridding procedure with a predetermined size. Subsequently, the entire imaging dataset was transformed into a tabular format, taking into account the spatial information of each cell. We utilized four different metrics to capture both local and global features in a frame including m1, m2 for both AP and DV axes, Ripley's k-function, and n (total number of cells in each grid). Here, m1 and m2 denote the first and second moments, respectively, capturing the distribution of active cells at each stage. Furthermore, Ripley's k-function was employed to analyze spatial correlation and quantify deviations from a random spatial distribution. Equation \ref{eq:1} illustrates the formula for calculating Ripley's k-function. Where, \textit{A} is the area under each window with constant radius, \textit {n} is the number of data points, $d_{ij}$ is the distance between two points, and  $e_{ij}$ is an edge correction weight.  Then, the tabular data went through two steps of averaging on each stage and time correcting. Since our goal is to predict the distribution of active cells in each stage and we have different numbers of frames for each stage, we averaged the whole feature values based on each stage. Also, to account for temporal alignment, we implemented a one-stage shift in features, where we utilized the features from the previous stage in prediction of the current stage.

\subsection{Training}
\label{subsec: intro}
Following the completion of the feature extraction process, the dataset undergoes preparation for training an XGBoost model, a supervised learning algorithm. The outcome of this pipeline is the count of active cells within each grid at a given stage, determined by the features from the preceding stage.

\subsection{Evaluation}
\label{subsec: intro}
Subsequent to training the model, its performance is evaluated using test data. During testing, all pre-processing and feature extraction steps are replicated, and the pre-trained XGBoost model is employed to forecast the count of active cells for each grid across various stages.

        \begin{equation}
        \begin{aligned}
         \vspace{-15pt}
            \hat K_ r = \frac {A}{n(n-1)} \sum_{i=1}^{n}\sum_{i=1, j\neq i}^{n} 1(d_{ij}\leq r)e_{ij}
        \end{aligned}
        \label{eq:1}
         \vspace{-15pt}
        \end{equation}


        \begin{figure}
        \vspace*{-10pt}
            \centering
            \centerline{\includegraphics[width= 9.3cm, height=9.5cm]{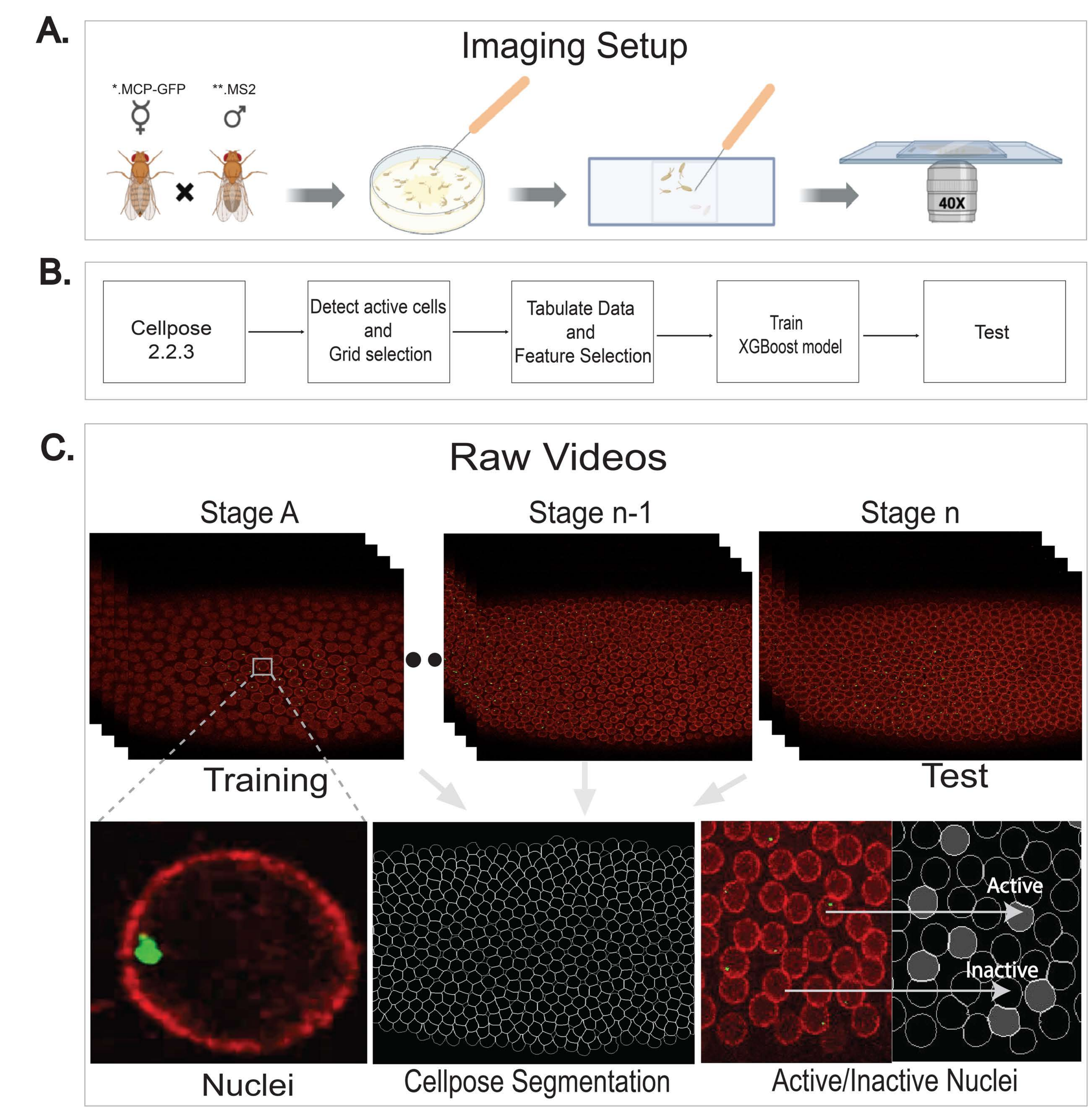}}
             \vspace{-5pt}
            \caption{Computational analysis of super-resolution live imaging compares nuclei activity and predicts stages. (A) Super-resolution live imaging set-up of hand-dechorionated Drosophila embryos of $MCP-GFP Nup-RFP (*.MCP-GFP) X  sogD\_ {\Delta Su(H)}-MS2$. (B) Implemented pipeline, starting with using Cellpose 2.2.3 for segmentation, followed by subsequent stages involving active celll detection, tabulating data and feature selection, training ,and testing. These steps collectively aim to predict the distribution of active cells for the next stage. (C) The $MCP-GFP-MS2$ system tracks transcription via $GFP-tagged$ $MCP$ binding to $MS2$ loops (Stage $A-$ NC13, double-dot ".." NC14$A$, NC14$B$, Stage n-1 NC14$C$)  and nuclei activity of live imaging snapshots is compared with Cellpose generated images.}
            \label{fig:2}
             \vspace{-5pt}
        \end{figure}



    \vspace{1cm}
\section{experiment and results}
\label{sec:intro}
\subsection{Comprehensive Analysis of super-resolution live movies } 
\label{subsec: main study}

  \begin{figure}
            \centering

            \centerline{\includegraphics[width= 9cm
            , height=8cm]{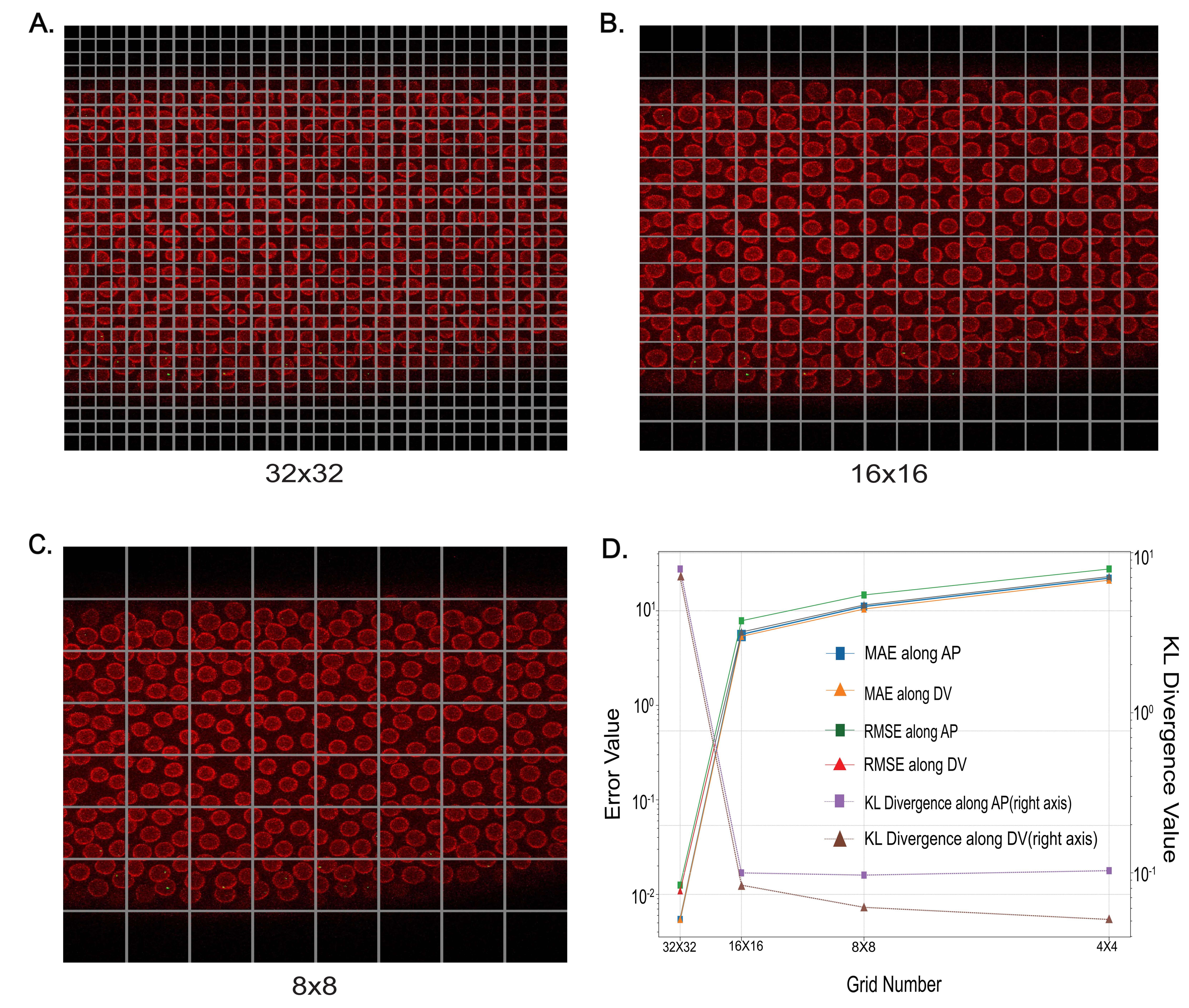}}
             \vspace{-10pt}
            \caption {Testing of optimal sampling parameters. (A–C) Depictions of three distinct grid configurations, labeled A, B, and C, corresponding to grid sizes of 32×32, 26×26, and 8×8, respectively. (D) presents the error plot associated with each grid configuration (A–C), facilitating the identification of the optimal grid size based on the lowest error value. }
            \label{fig:2}
             \vspace{-15pt}
        \end{figure}

\begin{figure}
            \centering
            \centerline{\includegraphics[width= 10cm, height=11cm]{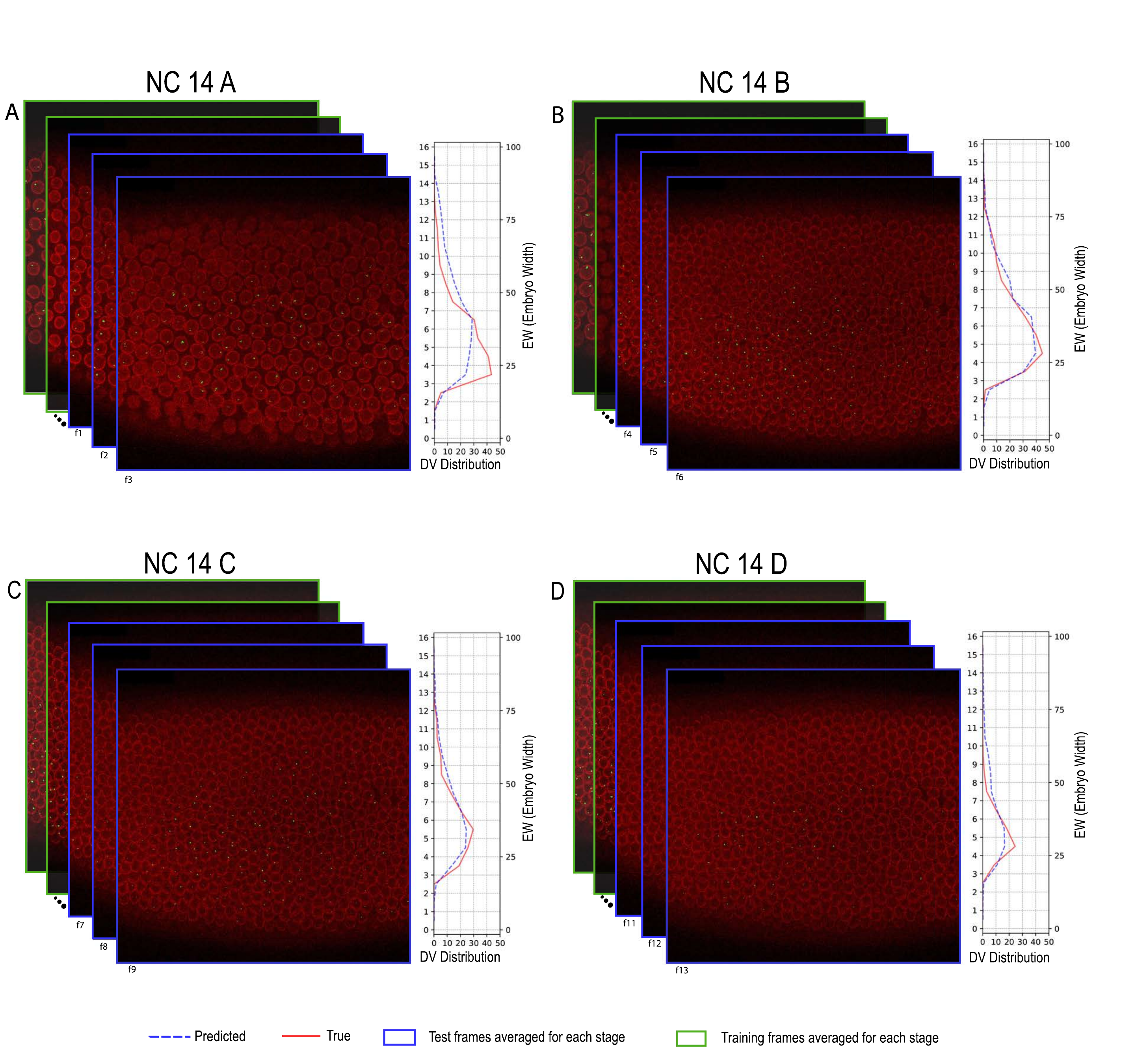}}
            \vspace{-10pt}
            \caption{The distribution of active cells achieving the best accuracy, based on mae values, is shown for the four stages of NC 14 (A–D). In panels A–D, green rectangles indicate the frames from the previous stage used to predict the blue frames of the current stage. The features from the previous stage frames were averaged to predict the average number of active cells in each grid for the current stage. For each stage, the right-hand plot illustrates the predicted and actual distribution of active cells along the DV axis, represented by dashed blue and red lines, respectively. In these plots, the grid numbers along the DV axis are shown from 0 to 16, the average number of active cells per grid is displayed from 0 to 50, and the embryo width along the DV axis spans from 0 to 100.}
            \label{fig:3}
            \vspace{-15pt}
        \end{figure}

 As outlined in the methodology section, during the feature extraction phase, square grids were applied to images, and the number of active cells within each grid was predicted. The key challenge was selecting the optimal grid size to enhance performance on test data. Consequently, we replicated the entire process of pre-processing and feature extraction for four distinct grid sizes: 250, 125, 62.5, and 31.25 (where the grid size of 'n' indicates the division of the entire image into \textit{n*n} squares). We used three different metrics to calculate the model performance on test data for different grid sizes which are rmse (root mean squared error), mae (mean absolute error), and Kullback-Leibler (KL) Divergence. Fig.\ref{fig:3} shows the experiment for different grid sizes. Our analysis revealed the same increasing trend in both rmse and mae as the grid size increases from 31.25 to 250 which indicated that a smaller grid size corresponds to a lower error. KL Divergence, which we also utilized as a metric, measures how one probability distribution diverges from a second one. Thus, the smaller value for it shows that two distributions are closer to each other. We used this criterion to see how well the pipeline can capture the trends in the active cells distribution. The KL Divergence for these four different grid sizes showed the different trend. Increasing the grid size from 31.25 to 250 yielded a decrease in KL Divergence. We had two options, the first one was to select 31.25 based on the lower rmse and mae. However, the problem was the average size of the cell was approximately 36 so if we set the grid size to 31.25 we have just one cell in each grid which changes the problem to a classification of active or inactive for each grid which was not our purpose. Another option was to select the optimal grid size based on KL Divergence, which finally, We selected the grid size of 62.5 over 31.25. The decision of selecting 63.5 over 125.0 although the 125 had lower KL Divergence, is attributed to the computational constraints of calculating Ripley's k-function for larger grid sizes in our setup.

In subsequent experiment, we conducted an ablation study to discern the relative importance of features, identifying those deemed crucial for inclusion in the final release and those that may be omitted. Table \ref{table:dataset} indicates the performance of different combinations of features. It can be concluded  that features of the first row including Ripley's k-function and n are the most important features that we used them for training and testing the pipeline. All reported mae values underwent the K-fold cross validation method to mitigate the influence of random results.


\begin{table} [htbp]
    \centering
    \begin{tabular}{|c|l|l|l|}
    \hline
        \textbf{Feature list} & \textbf{mae}  \\ 
        \hline
        n, Ripley's k-function & 3.799  \\ 
        \hline
        m2, n, Ripley's k-function & 3.86 \\
        \hline
       m2, m1\_AP, n, Ripley's k-function & 3.92 \\ 
        \hline
        Ripley's k-function  & 3.93 \\
        \hline
        m2, m1\_AP, m1\_DV, Ripley's k-function  & 3.94 \\
        \hline
    \end{tabular}
     \vspace{-5pt}
    \caption{The average mae value on K-fold cross validation over test dataset for different combinations of features for ablation study. }
    \vspace{-15pt}
    \label{table:dataset}
\end{table}

To visualize the performance of the pipeline with selected features and parameters, we tested the pre-trained model on test dataset. Fig \ref{fig:4} shows the active cell distribution for the best prediction based on the average mae values.

\subsection{Comparative Evaluation of $sogD$ and $sogD\_ {\Delta Su(H)}$}
\label{subsec:case and control}
  As, we had 6 videos for $sogD\_ {\Delta Su(H)}$ and 7 for $sogD$, we randomly selected 3 videos from each group for training and 1 for testing. Then, we averaged the AP\_mae, DV\_mae, and mean\_mae for whole $sogD\_ {\Delta Su(H)}$ and $sogD$ experiments and calculated the difference between $sogD\_  {\Delta Su(H)}$ and $sogD$ $sogD\_ {\Delta Su(H)}$ - $sogD$) for each of these metrics and the results were 0.210, 1.511, and 0.86 respectively. We also used cross-validation to avoid overfitting. These results show there is a difference between the performance of our pipeline on $sogD\_ {\Delta Su(H)}$ and $sogD$ in AP\_mean,  mean\_mae and DV\_mean. In other words, our method works better in predicting along the AP axis, the mean of AP, and DV on the $sogD$ data compared to the $sogD\_ {\Delta Su(H)}$ one. In order to substantiate this assertion, we conducted two additional experiments:

First, we leveraged Mixed-Effects modelling, which can account for both fixed effects (like the group: $sogD\_ {\Delta Su(H)}$ or $sogD$ and random effects (like the variation within videos and stages). The mixed-effects model can help in understanding the influence of these fixed and random effects on our dependent variables like DV\_mae, AP\_mae, mean\_mae. The goal is to understand whether there is a significant difference in any metrics between the $sogD\_ {\Delta Su(H)}$ and $sogD$ groups, accounting for the variability introduced by different stages. The $sogD$ has, on average, a lower AP\_mae compared to the $sogD\_ {\Delta Su(H)}$ by about 0.310 units with the P\_value of 0.476. It shows based on this test, there is not a statistically significant difference in AP\_mae between $sogD\_ {\Delta Su(H)}$ and $sogD$ groups. However, the result for DV\_mae shows the $sogD$ has lower value by 1.620 units and 0.001 P\_value. Also, the result for mean\_mae indicates $sogD$ has lower value by 0.971 units and 0.019 P\_value. Two latter results for DV\_mae and mean\_mae  indicate significant difference between $sogD-{\Delta Su(H)}$ and $sogD$. In other words, our performance on the $sogD$ outperforms $sogD\_ {\Delta Su(H)}$ one based on DV\_mae and mean\_mae.

\begin{figure*}
            \centering
            \centerline{\includegraphics[width=13cm,
            height=18cm]{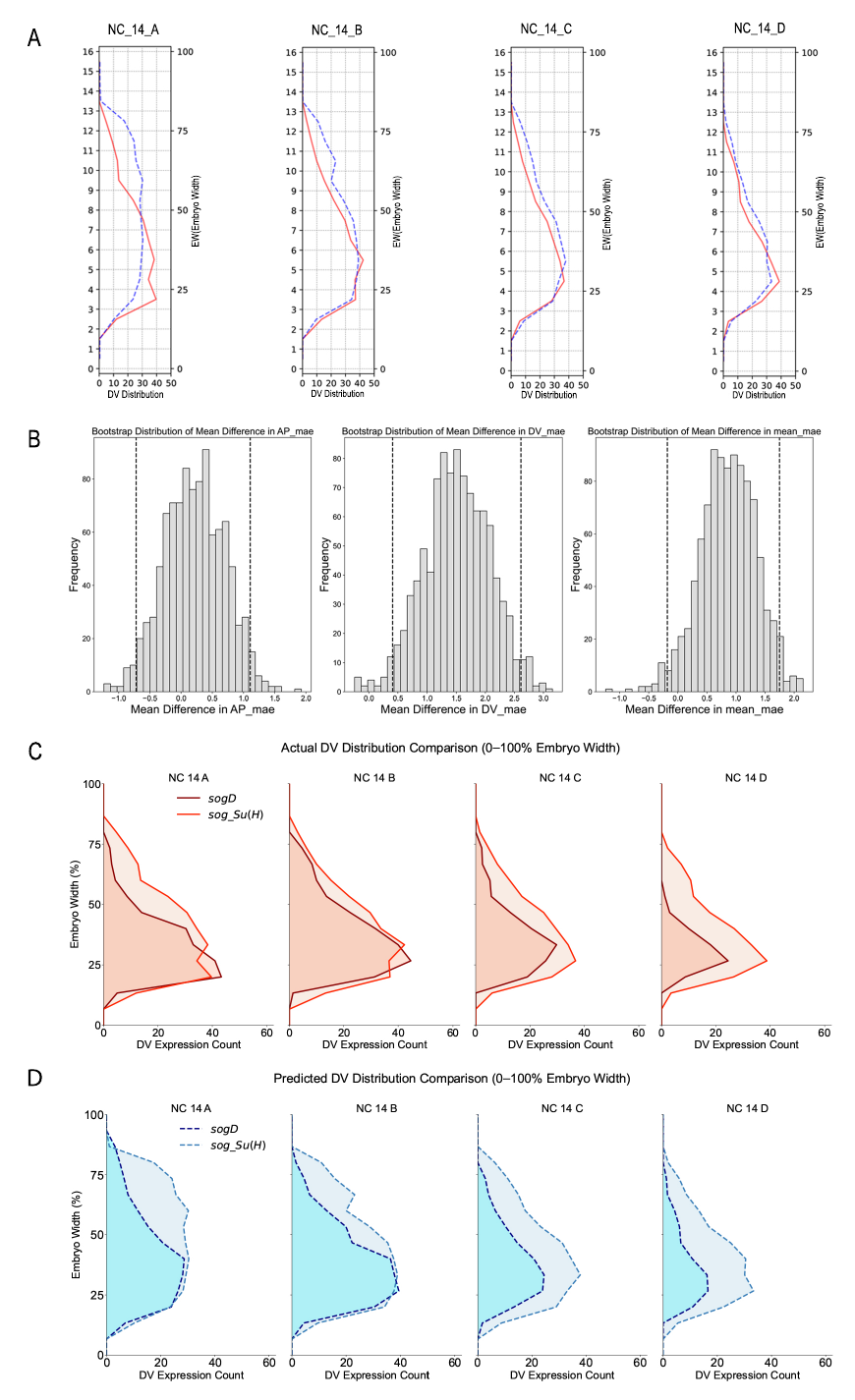}}
            \vspace{-15pt}
            \caption{(A) Distribution of active cells along the DV axis for the case dataset, where the red line represents the actual distribution and the dashed blue line corresponds to the predicted distribution. (B) Bootstrap distribution results for $AP-mae$ , $DV-mae$, and $mean-mae$ presented from left to right, respectively. (C) Actual DV distribution for the case and control datasets, shown in light and dark red, respectively, to illustrate changes in width over time.
            (D) Predicted DV distribution for the case and control datasets, represented in dashed light and dark blue, respectively.}
            \label{fig:4}
            \vspace{-15pt}
        \end{figure*}

In addition, we implemented another empirical hypothesis testing called the Bootstrap method. Bootstrap methods can be used to estimate the distribution of our metrics under the null hypothesis. To implement the bootstrap, we used the same metrics as previous method. we drew samples from the original dataset with replacement, to create a new dataset. Then, for each bootstrap sample, we computed the statistics of interest which are DV\_mae, AP\_mae, and mean\_mae. By analyzing the bootstrap distribution we can find the confidence intervals for each metrics. Fig. \ref{fig:4} -B shows the bootstrap distribution of mean difference in AP\_mae, DV\_mae, and mean\_mae. It indicates that with 95\% confidence interval the mean difference of DV\_mae, (DV\_mae($sogD\_ {\Delta Su(H)}$) - DV\_mae($sogD$)) was between [0.409 2.61]. It can be concluded that with 95\% confidence interval the DV\_mae for $sogD\_ {\Delta Su(H)}$ is at least 0.409 units higher than $sogD$, which means the performance of the pipeline for $sogD$ outperforms $sogD\_ {\Delta Su(H)}$ one. These ranges for AP\_mae and mean\_mae are respectively, [-0.72  1.10] and [-0.18  1.74 ]. 
It can be seen that for AP\_mae and mean\_mae the ranges include zero means the performance of $sogD$ can be better, equal, or worse than $sogD\_ {\Delta Su(H)}$. The results with Bootstrap method confirms the results derived from mixed effects method, which makes sense given that large amounts of training data are needed to model transgenic effects. 

    \section{Discussion}
\label{sec:disc}

In this study, we aimed to investigate whether a machine learning model can be trained to capture the trend of active cells during the embryogenesis process of \textit{Drosophila}. Following a comprehensive ablation study, we identified the optimal model architecture, feature set, and grid size configuration. We then evaluated our pre-trained model on test data from both the $sogD\_ {\Delta Su(H)}$ and $sogD$ datasets. Figure \ref{fig:3} presents the model’s predictions on the $sogD$ dataset, demonstrating strong performance in capturing the distribution of active cells along the DV axis, which is the primary focus of our study.  

Furthermore, we conducted an experiment to compare the model’s performance on the $sogD\_ {\Delta Su(H)}$ and $sogD$ datasets. The results indicate that the model performs better on the $sogD$ dataset, as reflected by lower DV\_MAE values. However, despite the overall superior performance on the $sogD$ dataset, the model successfully detected differences between the $sogD\_ {\Delta Su(H)}$ and $sogD$ conditions, particularly in NC 13C and NC 14D, where a reduction in embryo width of the active cell distribution along the DV axis was observed.

    \section{Conclusion}
Our work presents several key contributions. Firstly, we have developed a novel and optimized imaging technology that delivers spatial information throughout the entire DV axis of an embryo. Secondly, we introduce an automated pipeline that effectively discriminates cell types with high accuracy. Lastly, our approach enables the accurate prediction of the stage-level distribution of active cells, based on data from the preceding stage.
   %

   \section{Compliance with Ethical Standards}
All animal experiments were approved by the UTA IACUC review board. This study was performed in line with the principles of the Declaration of Helsinki. Approval was granted by the Ethics Committee of my institution. 

\section{Acknowledgments}
This work was supported by the Cancer Prevention and Research Institute of Texas (CPRIT) Recruitment of First-Time, Tenure-Track Faculty Members Grant (RR220015) (JML) and University of Texas System Rising STARs award (JML and TK). We would like to thank Mike Levine for providing fly lines and plasmids. We are also grateful to Angela Stathopoulos for generously supplying us with fly lines. We also thank all members of the Luber Lab for their productive discussions and the
Koromila Lab for their help with administrative tasks and fly husbandry.
    
    
    \bibliographystyle{IEEEbib}
    \bibliography{ref}

\begin{thebibliography}{10}

\bibitem{koromila2019distinct}
Theodora Koromila and Angelike Stathopoulos,
\newblock ``Distinct roles of broadly expressed repressors support dynamic enhancer action and change in time,''
\newblock {\em Cell reports}, vol. 28, no. 4, pp. 855--863, 2019.

\bibitem{dunipace2013autoregulatory}
Leslie Dunipace, Abbie Saunders, Hilary~L Ashe, and Angelike Stathopoulos,
\newblock ``Autoregulatory feedback controls sequential action of cis-regulatory modules at the brinker locus,''
\newblock {\em Developmental cell}, vol. 26, no. 5, pp. 536--543, 2013.

\bibitem{long2016ever}
Hannah~K Long, Sara~L Prescott, and Joanna Wysocka,
\newblock ``Ever-changing landscapes: transcriptional enhancers in development and evolution,''
\newblock {\em Cell}, vol. 167, no. 5, pp. 1170--1187, 2016.

\bibitem{perry2012precision}
Michael~W Perry, Jacques~P Bothma, Ryan~D Luu, and Michael Levine,
\newblock ``Precision of hunchback expression in the drosophila embryo,''
\newblock {\em Current biology}, vol. 22, no. 23, pp. 2247--2252, 2012.

\bibitem{lim2018visualization}
Bomyi Lim, Tyler Heist, Michael Levine, and Takashi Fukaya,
\newblock ``Visualization of transvection in living drosophila embryos,''
\newblock {\em Molecular cell}, vol. 70, no. 2, pp. 287--296, 2018.

\bibitem{birnie2023precisely}
Anthony Birnie, Audrey Plat, Cemil Korkmaz, and Jacques~P Bothma,
\newblock ``Precisely timed regulation of enhancer activity defines the binary expression pattern of fushi tarazu in the drosophila embryo,''
\newblock {\em Current Biology}, 2023.

\bibitem{lucas2013live}
Tanguy Lucas, Teresa Ferraro, Baptiste Roelens, Jose De Las~Heras Chanes, Aleksandra~M Walczak, Mathieu Coppey, and Nathalie Dostatni,
\newblock ``Live imaging of bicoid-dependent transcription in drosophila embryos,''
\newblock {\em Current biology}, vol. 23, no. 21, pp. 2135--2139, 2013.

\bibitem{la2018rna}
Gioele La~Manno, Ruslan Soldatov, Amit Zeisel, Emelie Braun, Hannah Hochgerner, Viktor Petukhov, Katja Lidschreiber, Maria~E Kastriti, Peter L{\"o}nnerberg, Alessandro Furlan, et~al.,
\newblock ``Rna velocity of single cells,''
\newblock {\em Nature}, vol. 560, no. 7719, pp. 494--498, 2018.

\bibitem{dayao2023deriving}
Monica~T Dayao, Alexandro Trevino, Honesty Kim, Matthew Ruffalo, H~Blaize D’Angio, Ryan Preska, Umamaheswar Duvvuri, Aaron~T Mayer, and Ziv Bar-Joseph,
\newblock ``Deriving spatial features from in situ proteomics imaging to enhance cancer survival analysis,''
\newblock {\em Bioinformatics}, vol. 39, no. Supplement\_1, pp. i140--i148, 2023.

\bibitem{fenelon2024h}
Kelli~D Fenelon, Priyanshi Borad, Biraaj Rout, Parisa~Boodaghi Malidarreh, Mohammad~Sadegh Nasr, Jacob~M Luber, and Theodora Koromila,
\newblock ``Su (h) modulates enhancer transcriptional bursting in prelude to gastrulation,''
\newblock {\em Cells}, vol. 13, no. 21, pp. 1759, 2024.

\end{thebibliography}
    \renewcommand{\thefigure}{Sup. Fig-\arabic{figure}}
\captionsetup[figure]{labelformat=onlynumber, labelsep=colon, labelfont=bf}
\setcounter{figure}{0}
\clearpage 
\onecolumn
\section{Supplemental Material}

\vspace{2pt}
 
\begin{figure*}[!ht]  
    \centering
    \includegraphics[width=9cm, height=3cm]{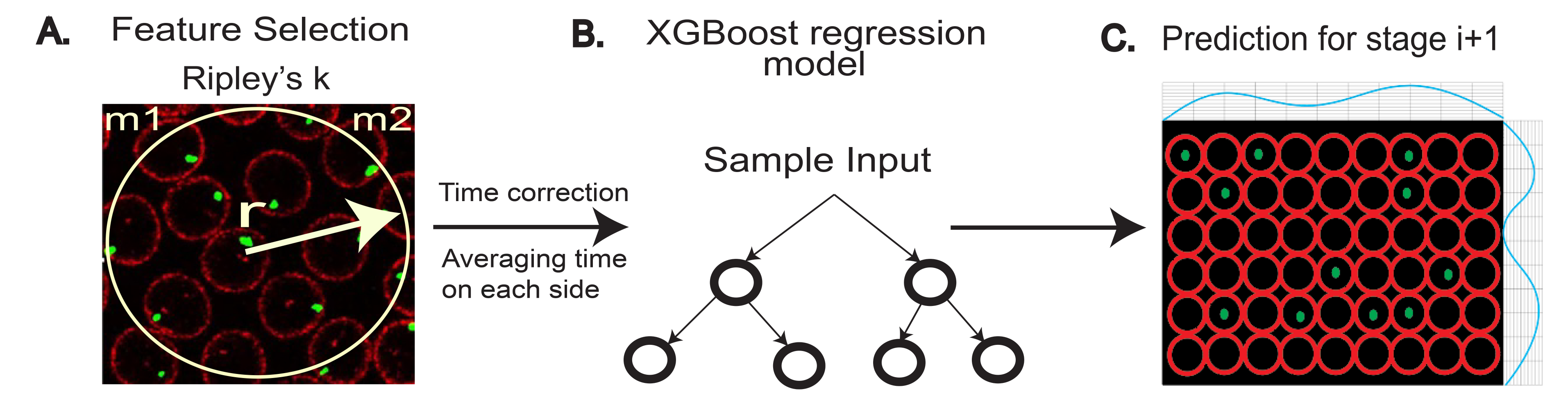}
    \vspace{-8pt}
    \caption{Detailed information about our pipeline, (A) Ripley's K function are used as a feature, (B) XGBoost is the machine learning method we used (C) the output of the pipeline that shows the distribution of active cells.}
    \label{sup. 1}
    \vspace{-15pt}
\end{figure*}

\vspace{3cm}

 \begin{figure*}[!ht]
            \centering
            \centerline{\includegraphics[width= 13cm
            , height=10cm]{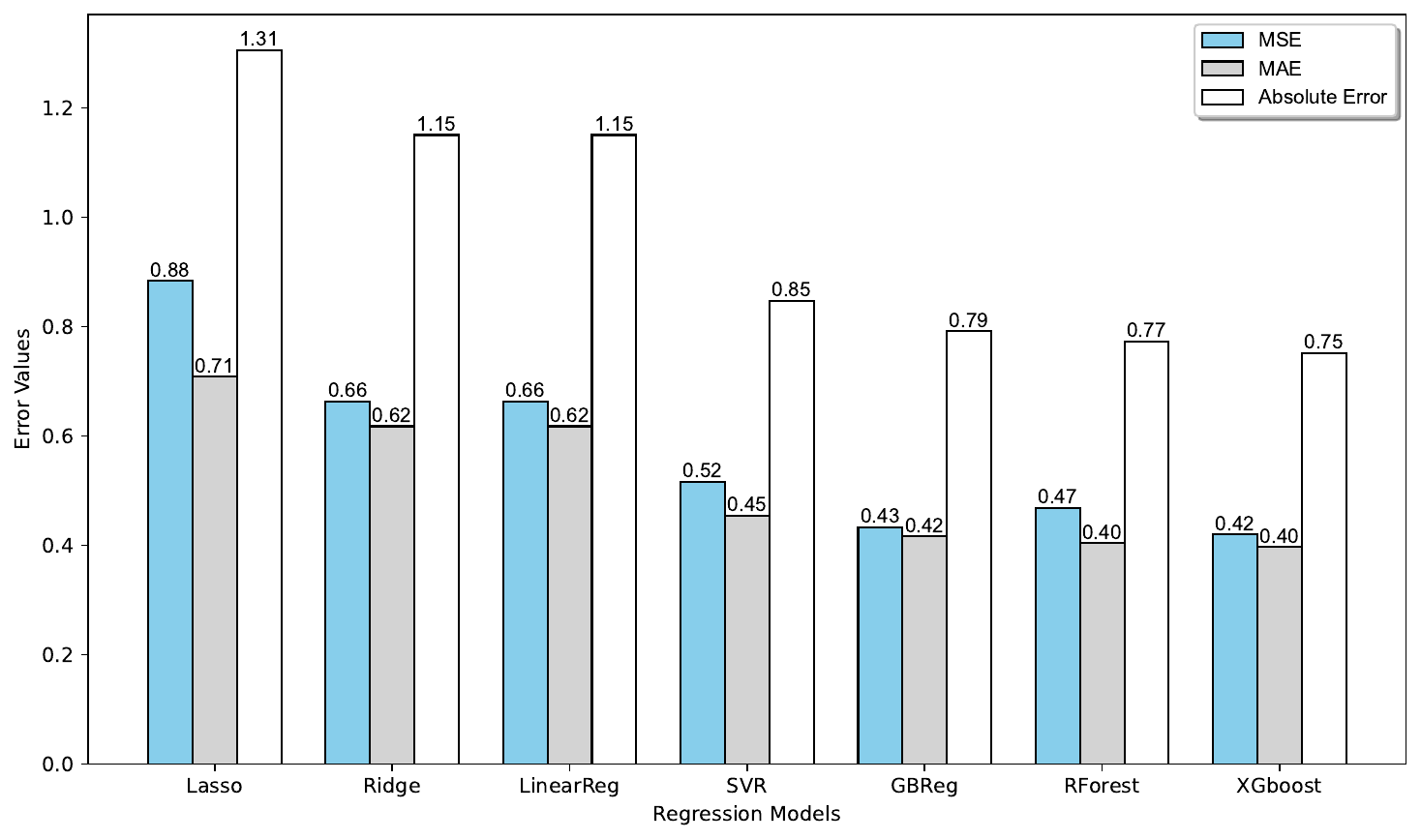}}
            \vspace{-15pt}
            \caption{ The comparison of different machine learning methods on control dataset based on three different metrics: MSE, MAE, and Absolute Error. It shows the best result corresponds to XGBoost model. }
            \label{supp2}
            \vspace{-15pt}
        \end{figure*}

\begin{figure*}
            \centering
            \centerline{\includegraphics[width= \linewidth
            , height=18cm]{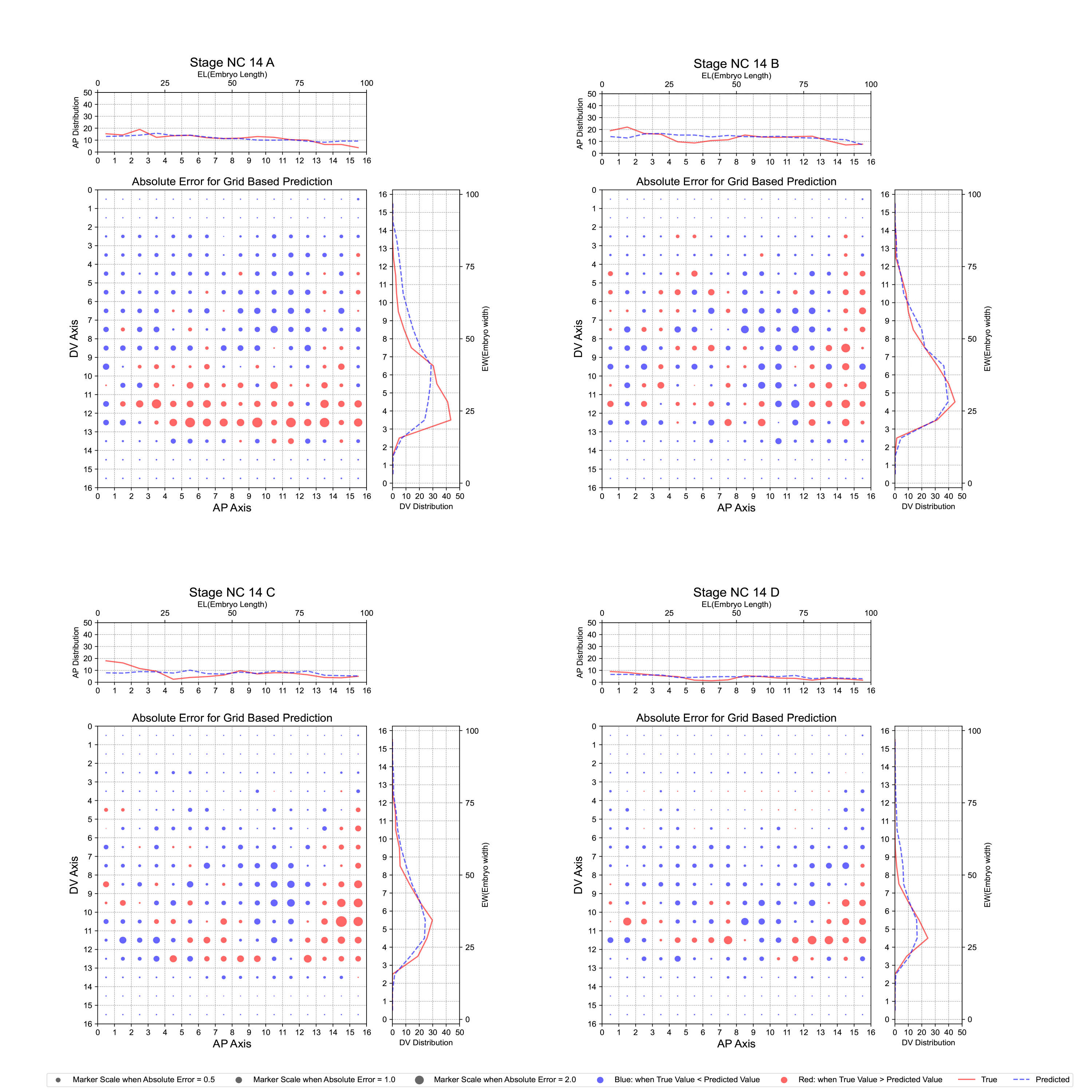}}
            \vspace{-8pt}
            \caption{ The distribution of active cell for the best  accuracy based on mae values for control (\textit{SogD}) dataset stages are NC 14 A-D. For each stage the top and right plot shows the distribution of active cells along AP and DV axis respectively. The middle plot shows the absolute error in each grid of the embryo to see where the prediction performs well. }
            \label{supp3}
            \vspace{-15pt}
        \end{figure*}

\begin{figure*}
            \centering
            \centerline{\includegraphics[width= \linewidth
            , height=18cm]{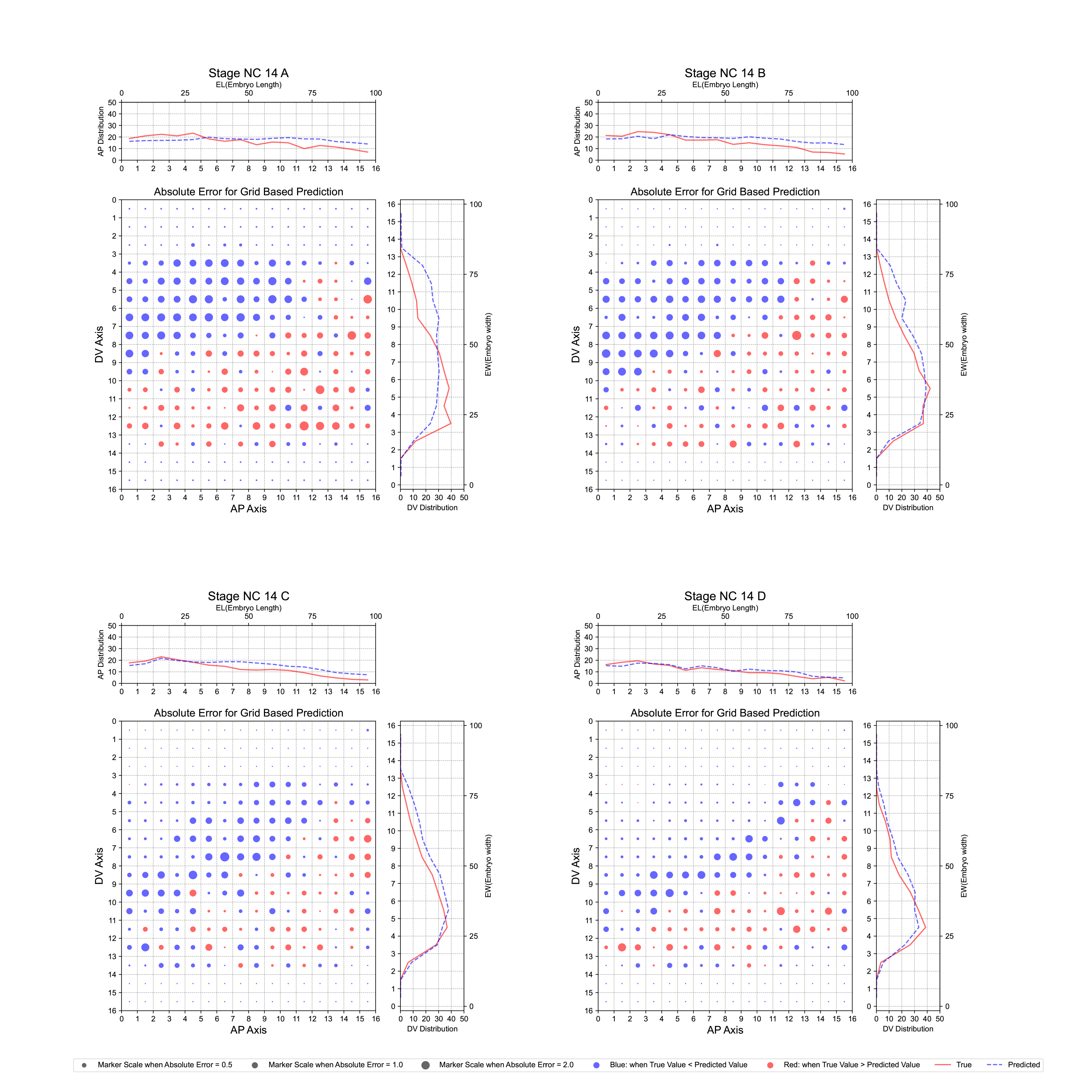}}
            \vspace{-15pt}
            \caption{ The distribution of active cell for the best  accuracy based on mae values for case ($sogD\_ {\Delta Su(H)}$) dataset stages are NC 14 A-D. For each stage the top and right plot shows the distribution of active cells along AP and DV axis respectively. The middle plot shows the absolute error in each grid of the embryo to see where the prediction demonstrates high reliability. }
            \label{supp4}
            \vspace{-15pt}
        \end{figure*}

\end{document}